\documentclass[aps,amsmath,amssymb,twocolumn,longbibliography,10pt,pra]{revtex4-1}
\usepackage[utf8]{inputenc}
\usepackage[T1]{fontenc}
\usepackage{graphicx}
\usepackage[colorlinks=true,linkcolor=blue,citecolor=red,urlcolor=magenta]{hyperref}

\begin{document} 

\title{Two Rydberg-dressed atoms escaping from an open well}
\author{Jacek Dobrzyniecki}
\email{Jacek.Dobrzyniecki@ifpan.edu.pl}
\author{Tomasz Sowi\'nski}
\affiliation{Institute of Physics, Polish Academy of Sciences, Aleja Lotnikow 32/46, PL-02668 Warsaw, Poland}

\date{\today} 

\begin{abstract}
A comprehensive analysis of the dynamics of two Rydberg-dressed particles (bosons or fermions) tunneling from a potential well into open space is provided. We show that the dominant decay mechanism switches from sequential tunneling to pair tunneling when the interaction strength is tuned below a certain critical value. These critical values can be modified by tuning the effective range of the interaction potential. By comparing the dynamics for bosons and fermions, we show that there are significant differences between the two cases. In particular, increasing the interaction range modifies the tunneling rate in opposite ways for fermions and bosons. Furthermore, for the fermionic system much stronger attractive interactions are needed to achieve pair tunneling. The results provide insight into the dynamics of tunneling systems and, in light of recent realizations of tunneling few-body systems and Rydberg dressing of atoms, they offer promise for future experiments. 
\end{abstract}

\maketitle

\section{Introduction}

The tunneling of particles from a potential well into empty space is one of the fundamental problems in quantum mechanics. It has been used in the analysis of such phenomena as the nuclear $\alpha$ decay \cite{1928-Gamow-ZP,1928-Gurney-Nature}, proton emission \cite{1999-Talou-PRC,2000-Talou-PRC}, fusion \cite{1998-Balantekin-RevModPhys}, fission \cite{1991-Bhandari-PRL}, photoassociation \cite{2000-Vatasescu-PRA}, photodissociation \cite{1984-Keller-PRA}, or the functioning of tunnel diodes \cite{1984-Ricco-PRB}. Many aspects of particle tunneling into open space have been studied in detail over the years. For example, the single-particle tunneling process and the tunneling of a multi-body Bose-Einstein condensate are now well understood \cite{1961-Winter-PR,2003-Razavy-Book,1998-Ueda-PRL,2001-Salasnich-PRA,2005-Carr-JPB,2006-Schlagheck-PRA,2007-Huhtamaki-PRA,2017-Zhao-PRA}. Between these two extreme situations lies the problem of tunneling of a few strongly interacting particles, which turns out to be a much more complicated issue. In this case, strong inter-body correlations play a role in the dynamics of the system, and thus the physics cannot be reduced to an approximate description at the one-body level \cite{2004-Gogolin-Book}. As a result, the problem of few-particle tunneling raises many questions that still have no satisfactory answers.

In recent years, interest in the subject of quantum tunneling has increased thanks to the rapid development of experimental techniques in the field of ultracold atom physics. It is possible to engineer systems with nearly any desired properties, such as the shape of the external potential \cite{2005-Meyrath-PRA,2009-Henderson-NJP,2010-VanEs-JPB}, the effective dimensionality \cite{2001-Gorlitz-PRL,2001-Greiner-PRL,2001-Schreck-PRL,2004-Stoferle-PRL}, the initial state \cite{2011-Serwane-Science}, or the strength of interparticle interactions \cite{2008-Pethick-Book,2010-Chin-RevModPhys}. Recent important experimental achievements in this area include the experiments in Selim Jochim's group in Heidelberg, where the decay of tunneling few-fermion systems was investigated \cite{2012-Zurn-PRL,2013-Zurn-PRL}.

The problem of a few particles tunneling from an open well has received significant attention in recent years, and multiple theoretical works on the subject have been published. In most of these works \cite{2006-DelCampo-PRA,2009-Lode-JPB,2011-Kim-JPB,2012-Maruyama-PRC,2012-Rontani-PRL,2012-Lode-PNAS,2013-Bugnion-PRA,2013-Hunn-PRA,2014-Lode-PRA,2013-Rontani-PRA,2014-Maksimov-PRA,2015-Gharashi-PRA,2015-Lundmark-PRA,2017-Ishmukhamedov-PRA,2019-Ishmukhamedov-PhysE,2020-Koscik-PRA}, the inter-particle interactions are assumed to be dominated by short-range forces, with only a few works \cite{2014-Krassovitskiy-JPB,2016-Fasshauer-PRA,2018-Oishi-JPG,2018-Oishi-Acta} focusing on long-range interactions. However, longer-range interacting systems can show interesting properties. There is a variety of approaches to creating long-range-interacting systems, such as e.g. using molecules or atoms with strong dipolar interactions \cite{2009-Lahaye-RepProgPhys}. 

One such possibility which has raised significant interest in recent years is the creation of cold atoms in so-called Rydberg-dressed states, which can be achieved when the ground atomic state is off-resonantly coupled to a high-lying Rydberg state \cite{2010-Pupillo-PRL,2010-Johnson-PRA,2010-Henkel-PRL,2010-Honer-PRL,2012-Li-PRA,2017-Plodzien-PRA}. Atoms in Rydberg-dressed states can exhibit strong interactions at large distances \cite{2016-Browaeys-JPB}, which at short distances saturate to a constant value \cite{2010-Johnson-PRA}. These interactions are highly controllable since the parameters of the interaction can be tuned by changing the parameters of the coupling laser. At the same time, Rydberg-dressed atoms avoid problems associated with ultra-cold atoms in bare Rydberg states, such as short lifetimes, or interaction energies large enough to overwhelm typical trapping potentials \cite{2010-Johnson-PRA}. Rydberg-dressed systems have been succesfully implemented in various setups, for both small and large systems \cite{2016-Jau-NatPhys,2016-Zeiher-NatPhys,2017-Zeiher-PRX,2019-Arias-PRL,2020-Borish-PRL}. They have many possible applications and can also be applied to systems in 1D geometry \cite{2017-Plodzien-PRA}. Recently, correlations in trapped two-atom systems with interactions of this kind were studied in \cite{2018-Koscik-SciRep,2019-Koscik-SciRep}. However, the correlations between two Rydberg-dressed particles tunneling from a potential trap have not yet been considered. 

In this paper we numerically analyze the dynamics of two particles (bosons or fermions) escaping from an effectively one-dimensional potential well into open space. The interaction potential is described by two freely tunable parameters: the approximate interaction range, and the effective interaction strength. We explore the dynamics of the particle tunneling for different interaction parameters and particle statistics. Similarly to our earlier studies of contact-interacting bosons \cite{2018-Dobrzyniecki-PRA,2019-Dobrzyniecki-PRA}, here we focus mainly on determining the dominant decay mechanism of the system: whether the particles tunnel sequentially (one by one), or as pairs. In this way we show how the tunneling dynamics of the system can be modified by tuning the interaction parameters. We additionally compare dynamical properties between bosonic and fermionic particles, showing how the quantum statistics affects the dynamical properties. It is worth mentioning that various aspects of pair tunneling in few-particle systems have been investigated previously \cite{2012-Maruyama-PRC,2013-Rontani-PRA,2015-Lundmark-PRA,2015-Gharashi-PRA,2018-Oishi-Acta,2018-Oishi-JPG,2019-Ishmukhamedov-PhysE}. However, in our work we give a comprehensive analysis of pair tunneling from different points of view, taking into account the interplay of quantum statistics, interaction strength and shape of interaction potential.

This work is organized as follows. In Sec.~\ref{sec:model} we describe the model system under study and the interaction potential. In Sec.~\ref{sec:initial-state} we examine the initial state of the system at $t = 0$, depending on the interaction parameters. In Sec.~\ref{sec:eigenstate-spectrum} we describe the spectrum of eigenstates of the two particles after opening the well. In Sec.~\ref{sec:short-time-dynamics} we describe the dynamics of the two-particle system, showing the basic nature of the tunneling dynamics, and the transition between distinct regimes that occurs at a specific value of the interaction strength. In Sec.~\ref{sec:long-time-dynamics} we focus on the long-time dynamics, analyzing the exponential nature of the decay. Section \ref{sec:conclusion} is the conclusion.

\section{The model}
\label{sec:model}

\begin{figure}[t]
\includegraphics[width=\linewidth]{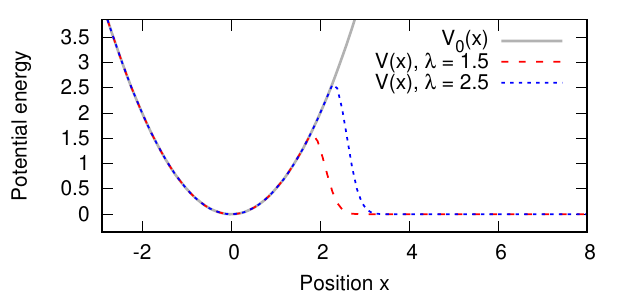}
\caption{The shape of the external potential at time $t < 0$ ($V_0(x)$, gray continuous line) and after the sudden change at $t = 0$ ($V(x)$, red and blue dotted line), for two different values of the parameter $\lambda = 1.5, 2.5$. Energy and length are shown in units of $\hbar\omega$ and $\sqrt{\hbar/m\omega}$, respectively.}
\label{fig:potential}
\end{figure}

We consider an effectively one-dimensional system of two identical spinless particles (bosons or fermions) of mass $m$, confined in an external potential $V(x)$ and interacting via the two-body interaction potential $U(r)$. The Hamiltonian of the system has the form 
\begin{equation}
H = \sum\limits_{i=1}^2 \left[ -\frac{\hbar^2}{2m} \frac{\partial^2}{\partial x_i^2} + V(x_i) \right] + U(x_1-x_2),
\end{equation}
where $x_i$ represents the position of the $i$-th particle. We assume that at time $t < 0$ the particles are confined inside a harmonic well potential with frequency $\omega$, $V_0(x) = \frac{1}{2} m \omega^2 x^2$. Then at $t = 0$ the well is opened from one side, and the external potential for $t \ge 0$ is given by
\begin{equation}
\label{eq:potential_t_gt_0}
 V(x) = 
 \begin{cases}
    \frac{1}{2}m\omega^2x^2 ,& x < \sqrt{2 \lambda}x_0, \\   
    \frac{1}{2}m\omega^2x^2 e^{-6(x/x_0-\sqrt{2 \lambda})^2} ,& x \ge \sqrt{2 \lambda}x_0,
\end{cases}
\end{equation}
where $x_0 = \sqrt{\hbar/m\omega}$ is the initial oscillator length unit. This potential has the form of a well separated from open space by a finite barrier and it is parametrized by the dimensionless parameter $\lambda$, approximately equal to the height of the barrier in units of $\hbar \omega$. The external potential $V(x)$ is shown in Fig.~\ref{fig:potential} and compared to the harmonic oscillator potential $V_0(x)$. The barrier height $\lambda$ is chosen so that it is higher than the energy of the system. This ensures that under-the-barrier tunneling is the only way to exit the well. For the bosonic system, we pick $\lambda=1.5$. In the case of fermions, we pick $\lambda=2.5$ since, due to the fermionic statistics, the energy of the initial state is different and a higher barrier height is necessary. 

\begin{figure}[t]
\includegraphics[width=\linewidth]{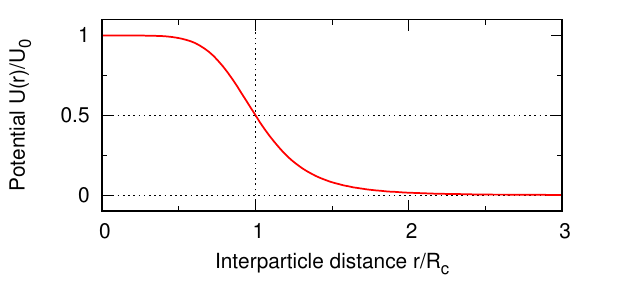}
\caption{The effective interaction potential $U(r)$ \eqref{eq:rydberg_interaction} as a function of interparticle distance $r$. The distance is expressed in terms of the effective range $R_\mathrm{c}$, and the potential energy is expressed in terms of the interaction amplitude $U_0$. At large distances the potential decays as $r^{-6}$, while at small distances ($|r| \lesssim R_\mathrm{c}$) it saturates to the constant value $U_0$.}
\label{fig:rydberg_interaction_potential}
\end{figure}

We assume the particles interact through a non-zero-range potential $U(r)$, in contrast to the contact interaction $g \delta(r)$ which is typically used to model interactions in ultracold systems. The form of the interaction potential $U(r)$ is based on the interaction between cold atoms in ``Rydberg-dressed'' states \cite{2010-Pupillo-PRL,2010-Johnson-PRA,2010-Henkel-PRL,2010-Honer-PRL,2012-Li-PRA,2016-Jau-NatPhys,2016-Zeiher-NatPhys,2017-Plodzien-PRA}. Experimentally, Rydberg dressing can be achieved by means of an off-resonant laser coupling between the atomic ground state and a highly-excited Rydberg state. As a result of this coupling, the ground state gains a small admixture of the Rybderg state. The effective interaction potential between such Rydberg-dressed atoms has a very characteristic form \cite{2010-Johnson-PRA,2010-Henkel-PRL,2010-Honer-PRL,2019-Koscik-SciRep}. At long interparticle distances, the interaction potential resembles the interaction between Rydberg atoms. We assume that in the studied case, the dominant contribution to this interaction are van der Waals forces that depend on the interatomic distance $r$ as $r^{-6}$. For short interparticle distances (below a certain critical range $R_\mathrm{c}$), the so-called Rydberg blockade effect suppresses a simultaneous excitation of two atoms, so the effective interaction saturates to a constant value as $r \to 0$ \cite{2010-Johnson-PRA}. The resulting effective interaction potential (under the assumption that the spatial size of the system in the perpendicular direction is much smaller than the interaction range) is modelled by the function \cite{2010-Honer-PRL,2010-Henkel-PRL,2012-Li-PRA,2016-Zeiher-NatPhys,2017-Plodzien-PRA}
\begin{equation}
\label{eq:rydberg_interaction}
 U(r) = U_0 \left[1+\left( \frac{r}{R_\mathrm{c}} \right)^6 \right]^{-1},
\end{equation}
where $U_0$ (having units of energy) is the interaction amplitude at $r = 0$, and $R_\mathrm{c}$ (having units of length) can be treated as the effective range of the interaction. Both these parameters can be independently regulated experimentally, being dependent on the detuning and the Rabi frequency of the coupling laser \cite{2010-Johnson-PRA,2012-Li-PRA,2016-Zeiher-NatPhys,2017-Plodzien-PRA}. The interaction potential \eqref{eq:rydberg_interaction} as a function of the interparticle distance is shown in Fig.~\ref{fig:rydberg_interaction_potential}. 

It is worth noting that in the limit $R_\mathrm{c} \to 0$, the interaction potential \eqref{eq:rydberg_interaction} is approximately equivalent to a contact interaction potential $g\delta(r)$ with $g = 2 R_\mathrm{c} U_0$ \cite{2018-Koscik-SciRep}. Basing on this fact, we adopt a convention that will allow us to compare the strength of interactions for different values of the range $R_\mathrm{c}$. Namely, we make the substitution $U_0 \to g/(2 R_\mathrm{c})$, and rewrite the potential \eqref{eq:rydberg_interaction} as
\begin{equation}
\label{eq:rydberg_interaction_rescaled}
 U(r) = \frac{g}{2 R_\mathrm{c}} \left[1+\left( \frac{r}{R_\mathrm{c}} \right)^6 \right]^{-1}.
\end{equation}

In this approach, the interaction is parametrized not directly by the amplitude $U_0$, but rather the effective interaction strength $g$ in the $R_\mathrm{c} \to 0$ limit. This convention has the benefit that it allows us to directly compare the bosonic system properties with those of a contact-interacting system, which have been previously analyzed e.g. in \cite{2018-Dobrzyniecki-PRA,2019-Dobrzyniecki-PRA} (although for a slightly different shape of the potential barrier). 

For convenience, in the following we express all magnitudes in natural units of the problem, i.e., energy is given in units of $\hbar\omega$, length in units of $\sqrt{\hbar/(m\omega)}$, interaction strength in units of $\sqrt{\hbar^3\omega/m}$, time in units of $1/\omega$, and momentum in units of $\sqrt{\hbar m \omega}$. 

As the system is initially confined in the harmonic oscillator trap $V_0(x)$, the initial two-body state of the system at $t = 0$ is taken to be the ground state of the interacting two-particle system confined in the potential $V_0(x)$. As there is no exact solution available for the case of interaction potential $U(r)$ (in contrast to the celebrated Busch \emph{et al.} solution for contact interactions \cite{1998-Busch-FoundPhys}), for given parameters $g$ and $R_\mathrm{c}$ we find the ground state numerically, by propagating a trial two-body wave function in imaginary time. The trial wave function is chosen as the ground state of two non-interacting particles (bosons or fermions) in a harmonic oscillator well.

The evolution of the system for $t > 0$ is calculated by integrating the time-dependent Schr\"{o}dinger equation numerically, using the fourth-order Runge-Kutta method with time step $\delta t = 0.005$. The calculations are done on a dense grid with spacing $\delta x = 0.125$, with the simulated region including a large extent of space in the region where the external potential vanishes. To clarify, we represent the two-body wave function $\Psi(x_1,x_2;t)$ by the amplitudes $\psi_{ij}(t)$, obtained after the decomposition $\Psi(x_1,x_2;t) = \sum_{ij} \psi_{ij}(t) [\varphi_i(x_1)\varphi_j(x_2) \pm \varphi_j(x_1)\varphi_i(x_2)]$. Here $\varphi_i(x)$ is a single-particle function being nonzero on the $i$-th grid-cell, \emph{i.e.}, $\varphi_i(x) = 1/\sqrt{\delta x}$ for $|x-x_i| \le \delta x / 2$. The extent of the simulated region is chosen as $x \in [-4,60]$, for a total of 512 grid points. To avoid reflections of the escaped particles off the boundary of the simulated region, we employ the complex absorbing potential technique \cite{1994-Riss-JPhysB,1996-Riss-JChemPhys,2004-Muga-PhysRep,2005-Shemer-PRA}. Specifically, in the region far from the trap (at $x > 30$) we add an imaginary potential term $-i\Gamma(x)$ to absorb particles. The form of the imaginary potential is chosen as the smoothly rising function $\Gamma(x) = 10^{-3} \times (x-30)^2$. We wish to emphasize we have carefully checked that the final results presented in the following do not depend on the details of $\Gamma(x)$. Details about the effects of the complex absorbing potential are available in appendix \ref{sec:cap-appendix} \footnote{The full Fortran simulation code is available from the authors upon request (Git commit hash: e4335a0ca754c1e73c64866eb2bf5f33cfbfe1d5)}. 

\section{Initial state and energy}
\label{sec:initial-state}

As noted, the initial state of the system is chosen as the ground state of two particles confined in a harmonic oscillator potential. As the properties of the initial state are directly connected to the subsequent evolution dynamics, we will now examine those properties in detail, depending on the interaction parameters. 

\subsection{Two-boson initial state}

\begin{figure*}[t]
\centering
\includegraphics[width=\linewidth]{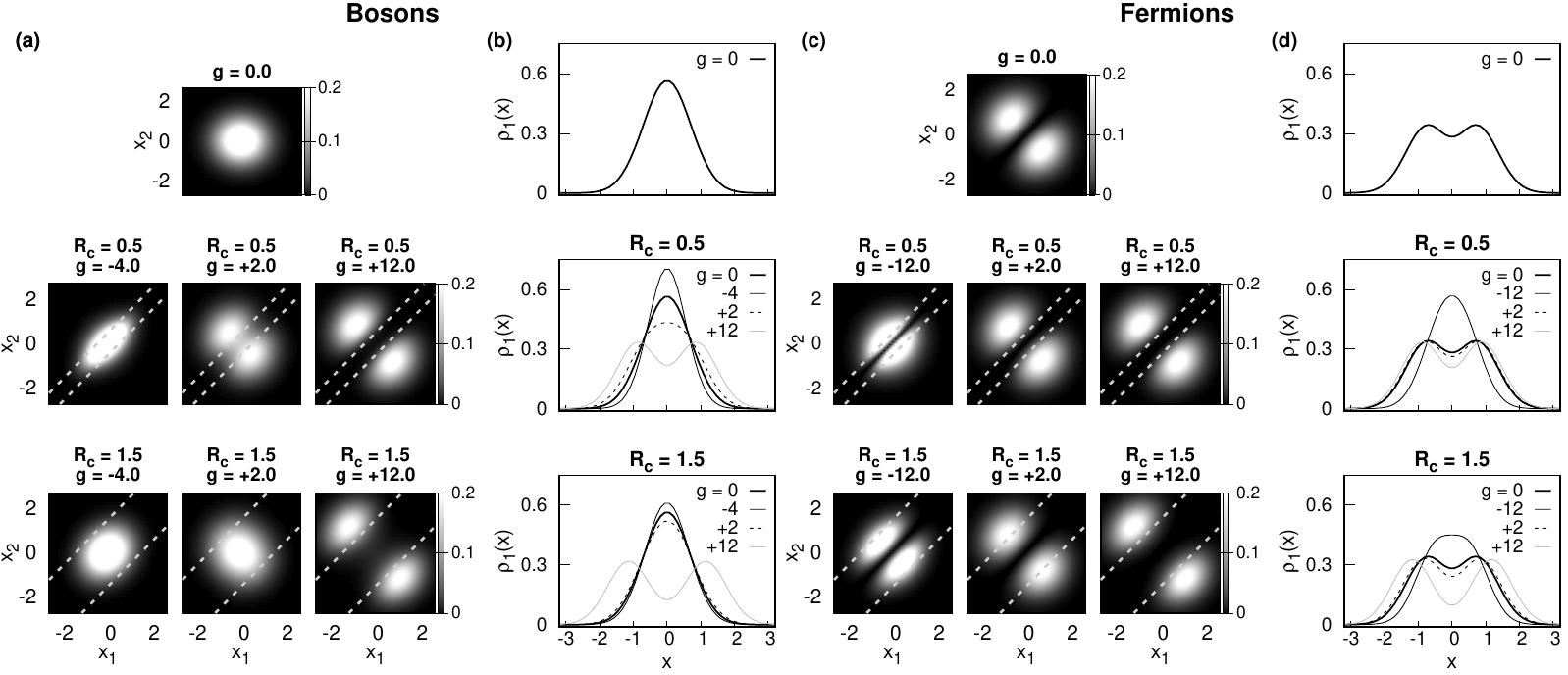}
\caption{(a) Two-body density distribution $\rho_2(x_1,x_2)$ of the initial state for the two-boson system, for varying values of interaction strength $g$ and interaction range $R_\mathrm{c}$. Gray dashed lines demarcate the region $|x_1-x_2| \le R_\mathrm{c}$, in which the distance between the bosons is within the interaction range. (b) The corresponding one-body density distribution $\rho_1(x)$ of the initial two-boson state, for varying $g$ and $R_\mathrm{c}$. (c) Two-body density distribution $\rho_2(x_1,x_2)$ of the initial state for the two-fermion system. (d) One-body density distribution $\rho_1(x)$ of the initial two-fermion state.  Lengths and range $R_\mathrm{c}$ are shown in units of $\sqrt{\hbar/m\omega}$, interaction strength is shown in units of $\sqrt{\hbar^3\omega/m}$.}
\label{fig:InitialStateDensity}
\end{figure*}

We first focus on the bosonic case. We will now directly examine the spatial distribution of two bosons in the initial state, to observe the relationship between interactions and particle correlations. In Fig.~\ref{fig:InitialStateDensity}a,b we show the single- and two-body density profiles [$\rho_1(x) = \int \mathrm{d}x |\Psi(x,x_2)|^2$ and $\rho_2(x_1,x_2) = |\Psi(x_1,x_2)|^2$] of the initial state for two bosons, for different interaction parameters. For clarity, the gray dashed lines in the $\rho_2$ plot indicate the boundaries of the two-body configuration space region for which $|x_1-x_2| \le R_\mathrm{c}$, \emph{i.e.}, the distance between the bosons is less than $R_\mathrm{c}$. For the non-interacting case ($g = 0$), both bosons are in the harmonic oscillator ground state, and thus both the two-particle density profile $\rho_2$ and the one-particle density profile $\rho_1$ have Gaussian shapes. In this case the boson positions are entirely uncorrelated with each other, and the two-body wave function is simply a product of two identical one-body wave functions. 

In the case of attractive interactions ($g = -4$), the boson positions become correlated. As can be seen from the profile $\rho_2$, the density becomes concentrated around the diagonal $x_1=x_2$, so that the bosons are more likely to be near each other. The attractive interactions also cause a narrowing of the one-body profile $\rho_1$, so the bosons are more likely to be found near the center of the well. However, for larger interaction range $R_\mathrm{c}$, the non-interacting wave function is already nearly completely contained within the region $|x_1-x_2| \le R_\mathrm{c}$ and within that region the felt interaction is nearly constant. As a result, for higher $R_\mathrm{c}$ the attractive interactions do not significantly change the shape of the density profile. 

For repulsive interactions ($g = +2$ and $g = +12$), bosons are less likely to be found near each other. For large enough interaction strength the two-body density in the region $|x_1-x_2| \le R_\mathrm{c}$ is nearly completely depleted, and the density profile $\rho_1$ splits into two maxima away from each other, indicating that the bosons are likely to be found on the opposite sides of the well. For large interaction ranges $R_\mathrm{c}$ the effect of the repulsions on the density profile is weakened, so that a larger repulsive interaction strength is needed to empty the region $|x_1-x_2| \le R_\mathrm{c}$. This is because, as $R_\mathrm{c}$ increases, pushing the bosons away from each other towards the well edges requires a higher energy cost. 

\begin{figure*}[t]
\centering
\includegraphics[width=\linewidth]{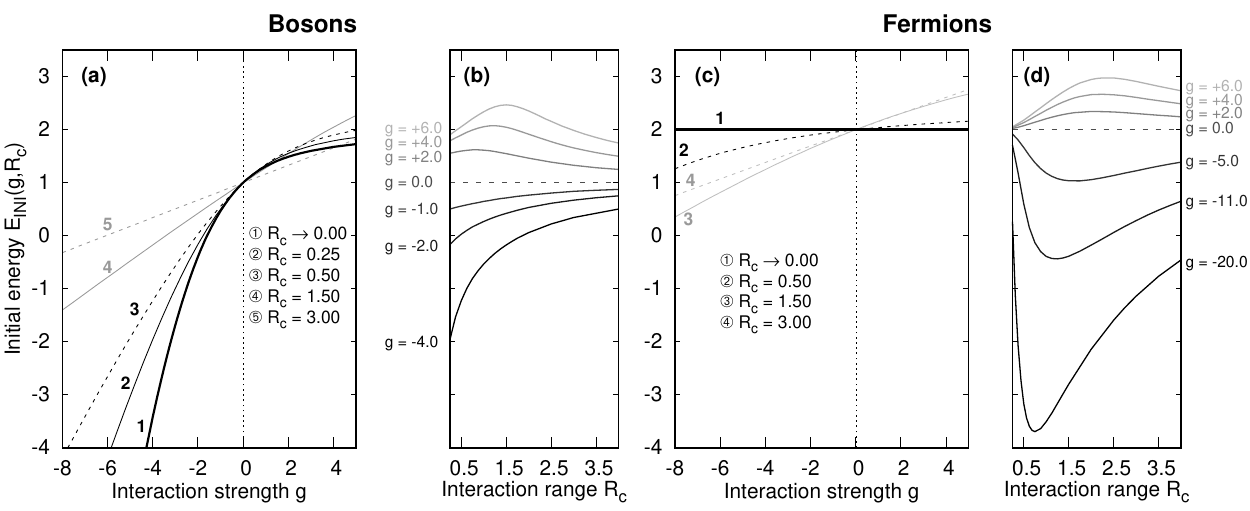}
\caption{(a) Initial state energy $E_\mathrm{INI}(g,R_\mathrm{c})$ for the two-boson system as a function of the interaction strength $g$, for different interaction ranges $R_\mathrm{c}$. (b) The energy $E_\mathrm{INI}(g,R_\mathrm{c})$ of two bosons as a function of $R_\mathrm{c}$, with $g$ constant. (c) Initial state energy $E_\mathrm{INI}(g,R_\mathrm{c})$ for the two-fermion system, as a function of the interaction strength $g$. Note that $E_\mathrm{INI}$ is a non-monotonic function of $R_c$, as shown in the next subfigure. (d) The energy $E_\mathrm{INI}(g,R_\mathrm{c})$ of two fermions as a function of $R_\mathrm{c}$. Energy is given in units of $\hbar \omega$, range $R_\mathrm{c}$ in units of $\sqrt{\hbar/m\omega}$, interaction strength $g$ in units of $\sqrt{\hbar^3\omega/m}$.}
\label{fig:InitialEnergy}
\end{figure*}

Now let us analyze the initial energy of the two-boson system. In Fig.~\ref{fig:InitialEnergy}a we show the energy $E_\mathrm{INI}(g,R_\mathrm{c})$ of two bosons for different interaction strengths $g$ and interaction ranges $R_\mathrm{c}$. Also shown is the energy calculated in the contact interaction limit $R_\mathrm{c} \to 0$, \emph{i.e.}, for bosons interacting via the contact potential $g \delta(r)$. The energy is calculated for a system in the harmonic oscillator potential $V_0(x)$, but after the external potential is changed to $V(x)$ at $t=0$, the energy of the system is almost unchanged (since the potential in the initial confinement region remains almost the same). 

In the $R_\mathrm{c} \to 0$ limit, the energy is a monotonic function of $g$. As the interaction range $R_\mathrm{c}$ increases, the energy becomes overall less sensitive to changes in the interaction strength (the slope of $E_\mathrm{INI}(g,R_\mathrm{c})$ measured at $g = 0$ becomes smaller). Although in this work we focus only on interaction ranges $R_\mathrm{c} \le 1.5$ (on the order of a single natural length unit), it should be pointed out that in the $R_\mathrm{c} \to \infty$ limit the interaction $U(r)$ is expected to vanish completely for all finite $g$. This is because for $R_\mathrm{c}$ approaching infinity, the interaction is felt simply as an energy shift constant in space, with magnitude $g/(2R_\mathrm{c})$. When $R_\mathrm{c} \to \infty$, this energy shift goes to zero for all finite $g$. 

To better understand the effect of the interaction range, in Fig.~\ref{fig:InitialEnergy}b we examine the dependency of $E_\mathrm{INI}(g,R_\mathrm{c})$ on $R_\mathrm{c}$, with fixed $g$. For attractive interactions, the energy has a monotonic dependency on $R_\mathrm{c}$ and gradually approaches the non-interacting value as $R_\mathrm{c}$ increases. This agrees with the previously observed properties of the density profile: for increasing $R_\mathrm{c}$, the density profile is less squeezed and smoothly approaches the non-interacting profile. On the other hand, for repulsive interactions, the dependency of $E_\mathrm{INI}(g,R_\mathrm{c})$ on $R_\mathrm{c}$ is not monotonic. For smaller $R_\mathrm{c}$, the energy increases with $R_\mathrm{c}$ until a certain maximum value, then it begins decreasing, approaching the non-interacting value $E_\mathrm{INI} = 1$. This observation can likewise be explained by considering the density profile for repulsive systems. At first, increasing $R_\mathrm{c}$ causes the bosons to be pushed away from each other towards further regions of the harmonic well, increasing the system energy. Beyond a certain interaction range, the interaction energy for a given $g$ is no longer sufficient to separate the bosons to a distance $\sim R_\mathrm{c}$, thus for high $R_\mathrm{c}$ the state density profile is identical to the non-interacting one. 

\subsection{Two-fermion initial state}

Let us now proceed to the two-fermion case. Owing to the different particle statistics, already on the level of the initial state this case differs visibly from the bosons. In Fig.~\ref{fig:InitialStateDensity}c,d we show the two-body and one-body density profiles $\rho_2$ and $\rho_1$ for the initial two-fermion state, at different interaction ranges $R_\mathrm{c}$ and interaction strengths $g$. As before, gray dashed lines in the $\rho_2$ plots indicate the region where $|x_1-x_2| \le R_\mathrm{c}$. In the non-interacting case ($g = 0$), the initial two-body state is the antisymmetrized product of the two lowest harmonic oscillator orbitals. As a result, the two-body density profile $\rho_2$ is entirely different from the bosonic case. The particle positions are anticorrelated, so that the fermions are more likely to be found on opposite sides of the well. The one-body density profile $\rho_1$ has a characteristic shape with two maxima located at opposite sides from the well center. The Pauli principle is manifested by the impossibility to find the two fermions at exactly the same position (\emph{i.e.}, the density along $x_1=x_2$ is empty). 

For attractive interactions ($g = -12$), the density is more concentrated within the $|x_1-x_2| \le R_\mathrm{c}$ region, \emph{i.e.}, the two fermions are more likely to be close to each other, although the $x_1 = x_2$ diagonal remains empty. Furthermore, for strong enough attractions, the two maxima in $\rho_1$ fuse into one maximum located in the center of the well. As $R_\mathrm{c}$ increases, the effect of attractions on the density profile becomes weaker, for the same reason as for bosons: for large $R_\mathrm{c}$ most of the entire non-interacting density profile is already contained within the $|x_1-x_2| \le R_\mathrm{c}$ region. 

In the case of repulsive interactions ($g = +2, g = +12$), another important difference compared to the boson case can be seen. Namely, for small interaction range ($R_\mathrm{c} = 0.5$), the density profile is almost unaffected by the repulsions. This is because the non-interacting two-body wave function already vanishes in such close vicinity to the diagonal, and any further repulsions do not modify it significantly. Only for higher interaction range ($R_\mathrm{c} = 1.5$) the density profiles are seen to be affected by the repulsive interactions, with the fermions pushed further away from each other. It is worth pointing out that for large $R_\mathrm{c}$ and $g$, both in the case of bosons and fermions there occurs a complete separation between the particles, and certain properties of the system (such as the density profile) become insensitive to the particle statistics in this case. 

We now turn our attention to the initial energy $E_\mathrm{INI}(g,R_\mathrm{c})$. In Fig.~\ref{fig:InitialEnergy}c we show the two-fermion $E_\mathrm{INI}(g,R_\mathrm{c})$ for different interaction parameters $g$ and $R_\mathrm{c}$. The vanishing of the two-fermion wave function at $r = 0$ means that the energy is overall less affected by interactions than in the bosonic case. In the limit $R_\mathrm{c} \to 0$, it furthermore means that the interaction $U(r)$ is not felt at all, and the energy in this case is independent of interactions: $E_\mathrm{INI}(R_\mathrm{c} = 0) = 2$. As $R_\mathrm{c}$ increases above zero, the energy gradually becomes more sensitive to interactions (as can be seen from the increasing slope of $E_\mathrm{INI}$ near the $g = 0$ point). Note that this is directly opposite to the boson case, where increasing $R_\mathrm{c}$ causes the energy near $g = 0$ to becomes less sensitive to interactions. However, it should be noted that in the $R_\mathrm{c} \to \infty$ limit the interaction is no longer felt by the two-fermion system, for the same reason as with bosons. Thus, for large enough $R_\mathrm{c}$ the trend reverses, at which point further increase of $R_\mathrm{c}$ causes the energy to approach the non-interacting value. 

For a clearer demonstration of how the two-fermion energy depends on the interaction range, in Fig.~\ref{fig:InitialEnergy}d we show the dependency of $E_\mathrm{INI}(g,R_\mathrm{c})$ on $R_\mathrm{c}$, with $g$ constant. The major difference from the bosonic case is that the energy approaches the same constant value in the two limits $R_\mathrm{c} \to 0$ and $R_\mathrm{c} \to \infty$. Thus, for intermediate values of $R_\mathrm{c}$ the energy has a non-monotonic dependency on $R_\mathrm{c}$, with a single minimum (maximum) for attractive (repulsive) interactions. 

\section{Eigenstates of two particles in open space}
\label{sec:eigenstate-spectrum}

After the well is opened at $t = 0$, the initial state starts to decay as the particles start tunneling into open space. To gain a basic understanding of the tunneling process, it is helpful to examine the many-body Hamiltonian spectrum for a system of particles in the region outside the well. In this way we can understand what configurations are available for the escaping particles. 

For this purpose, we describe the particles in their end-state (after tunneling) via a simplified Hamiltonian. We assume that the particles are far enough from the well that they feel no external potential, and thus can be described by a simplified Hamiltonian with $V(x) = 0$: 
\begin{equation}
\label{eq:open-space-hamiltonian}
H_\mathrm{out} = \sum\limits_{i=1}^2 \left[ -\frac{\hbar^2}{2m} \frac{\partial^2}{\partial x_i^2} \right] + U(x_1-x_2). 
\end{equation}

To find the eigenstates and eigenenergies of the Hamiltonian \eqref{eq:open-space-hamiltonian}, it is convenient to perform a transformation to the coordinates of the center-of-mass frame: $X = (x_1+x_2)/2, r = x_1 - x_2$. In these new variables the Hamiltonian can be written as a sum of two independent single-particle Hamiltonians, $H_\mathrm{out} = H_\mathrm{X} + H_\mathrm{r}$:
\begin{align}
H_\mathrm{X} &= -\frac{1}{4}\frac{\partial^2}{\partial X^2}, \label{eq:hamiltonian_rel_rydberg_cm} \\
H_\mathrm{r} &= -\frac{\partial^2}{\partial r^2} + U(r). \label{eq:hamiltonian_rel_rydberg_r}
\end{align}
The total energy of the two particles in free space is correspondingly a sum of eigenenergies of the two Hamiltonians, $E = E_\mathrm{X} + E_\mathrm{r}$, and the wave function is given in terms of the product of their eigenfunctions, $\Psi(x_1,x_2) = \phi_\mathrm{X}(X) \phi_\mathrm{r}(r)$. 

Solutions for the center-of-mass motion Hamiltonian $H_\mathrm{X}$ are straightforward, representing free-particle wave functions. In case of the relative-motion Hamiltonian $H_\mathrm{r}$, an exact solution is not available, and we obtain the eigenenergies and eigenfunctions by numerical diagonalization. 

\begin{figure}[t]
\centering
\includegraphics[width=\linewidth]{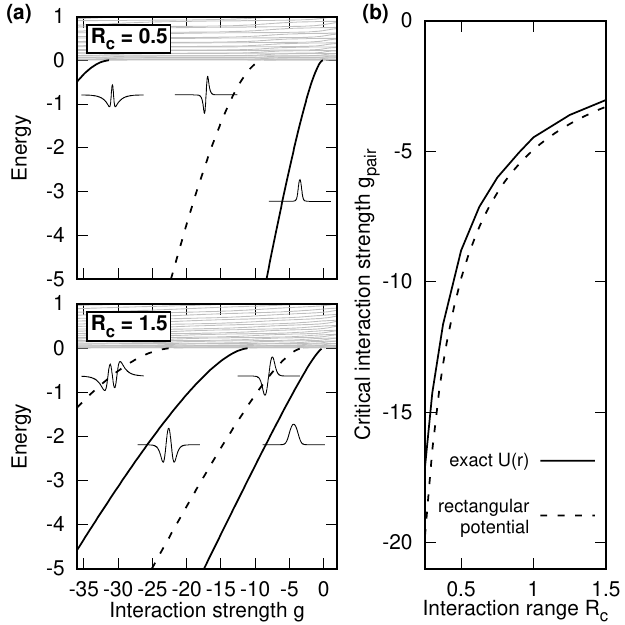}
\caption{(a) The two-body spectrum of the relative-motion Hamiltonian $H_\mathrm{r}$ \eqref{eq:hamiltonian_rel_rydberg_r} for two particles in empty space, interacting by the potential $U(r)$, as a function of interaction strength $g$. Results are shown for two different interaction ranges: $R_\mathrm{c} = 0.5$, $R_\mathrm{c} = 1.5$. For all $g$ there exists a spectrum of scattering states with $E_\mathrm{r} > 0$ (gray) that describe the relative motion of two almost-free particles. For $g < 0$ there are also bound states available, with energy $E_\mathrm{r} < 0$ (black). Solid (dashed) black lines correspond to bound states which have wave functions $\phi_\mathrm{r}(r)$ symmetric (antisymmetric) about $r=0$. The general shape of the wave functions $\phi_\mathrm{r}(r)$ is shown schematically near the corresponding energies. (b) The threshold interaction strength $g_\mathrm{pair}$, below which there exists an antisymmetric bound state in the $H_\mathrm{r}$ eigenspectrum and thus pairing of two fermions is possible. The shown results are those calculated numerically for the exact potential $U(r)$ (solid line), and the result $g_\mathrm{pair} \approx -\pi^2/(2 R_\mathrm{c})$ for an approximate rectangular potential (dashed line). Close agreement is seen between the two values. Energies are expressed in units of $\hbar\omega$, interaction strength in units of $\sqrt{\hbar^3\omega/m}$, interaction range in units of $\sqrt{\hbar/m\omega}$.}
\label{fig:BosonR-RydbergRelativeHamiltonianEigenspectrum}
\end{figure}

In Fig.~\ref{fig:BosonR-RydbergRelativeHamiltonianEigenspectrum}a we show the spectrum of eigenenergies of $H_\mathrm{r}$ as a function of $g$, obtained by numerical diagonalization, for two different values of $R_\mathrm{c}$. There are two groups of states distinguishable. The first group (indicated in gray) consists of almost-free-particle states with positive energy $E_\mathrm{r}$, forming a dense band. Their relative wave functions $\phi_\mathrm{r}(r)$ have a density distributed throughout all space, and describe a configuration of two (nearly) free particles. These states are present for all values of $g$. The second group (indicated in black) includes bound states with negative energy $E_\mathrm{r}$. They are much more sparse than the scattering states and do not form a dense band. Their wave functions $\phi_\mathrm{r}(r)$, with density centered near $r = 0$, describe states of two bound particles travelling together. These states only appear for negative interaction strengths $g < 0$. 

The wave functions $\phi_\mathrm{r}(r)$ have a well-defined symmetry in $r$, being even or odd functions of $r$: $\phi_\mathrm{r}(-r) = \pm \phi_\mathrm{r}(r)$. A single symmetric bound state appears immediately below $g = 0$ (indicated with a solid black line). For increasing attractive interactions $|g|$, additional bound states make their appearance, alternating between anti- and symmetric wave functions $\phi_\mathrm{r}(r)$ (their energies are indicated by dashed and solid black lines, respectively). The spacing between values of $g$ at which subsequent bound states appear is dependent on $R_\mathrm{c}$. For decreasing $R_\mathrm{c}$, the spacing between the bound states increases, and in the limit $R_\mathrm{c} \rightarrow 0$ (where the potential becomes equivalent to the contact potential) only one symmetric bound state is present. 

The possibility that the particles will be able to form pairs in the outside-well region depends on the availability of appropriate bound states. For bosons, where the relative wave function must be symmetric, the appropriate bound state becomes available as soon as interaction strength is below zero ($g < 0$), regardless of the value of $R_\mathrm{c}$. However, for fermions, the necessary bound state must have an antisymmetric wave function. Thus, pairing for fermions is only possible below a certain value $g_\mathrm{pair} < 0$, for which a second bound state (with odd symmetry) appears in the spectrum. This value $g_\mathrm{pair}$ is directly dependent on $R_\mathrm{c}$. 

It is worth noting that the approximate value of $g_\mathrm{pair}$ can be obtained analytically when the interaction potential $U(r)$ in \eqref{eq:hamiltonian_rel_rydberg_r} is replaced by a rectangular well potential, since in this case there exists an exact expression for the total number $n$ of bound states \cite{2003-Williams-Book}. For the particular parameters in this problem (mass $1/2$, well length $2R_\mathrm{c}$, well depth $|g|/(2R_\mathrm{c})$) the expression is $n = \lceil \sqrt{2 R_\mathrm{c} |g| } / \pi \rceil$, where $\lceil \cdot \rceil$ is the ceiling function, \emph{i.e.}, rounding up to the nearest integer. Therefore, the condition for the existence of a second bound state is $\sqrt{2 R_\mathrm{c} |g| } / \pi > 1$, giving the expression for $|g_\mathrm{pair}|$ as $\pi^2/(2 R_\mathrm{c})$. In Fig.~\ref{fig:BosonR-RydbergRelativeHamiltonianEigenspectrum}b we compare this expression with the numerically obtained value of $g_\mathrm{pair}$ for the Rydberg potential (defined as the highest value of $g$ at which there are at least two states with negative energy). We obtain a close agreement between the two cases. Note that in the limit of contact interactions ($R_\mathrm{c} \rightarrow 0$) we have $g_\mathrm{pair} \rightarrow -\infty$, so that the pairing between fermions becomes impossible, as expected for a contact potential limited to the $s$-wave scattering level.

The above results have direct significance for the tunneling dynamics. It can be surmised that the presence of pair tunneling depends on whether the particles are able to form pairs in the open-space region. The above analysis indicates that for bosons, pair tunneling will be present to some degree for any value of attractive interactions. For fermions, much greater interaction scales will be needed to analyze pair tunneling, since a strong attractive interaction $g < g_\mathrm{pair}$ is needed for pair tunneling to even occur in the first place. 

However, while this eigenspectrum gives information about the availability of specific states, it does not directly specify which of the tunneling mechanisms will dominate in the dynamics. We therefore address this question by performing a numerically exact time evolution and analyzing the tunneling process in a time-dependent way. 

\section{Dynamics of the density distribution}
\label{sec:short-time-dynamics}

The dynamics at $t>0$ can be quite well understood when the evolution of the two-body density distribution $\rho_2(x_1,x_2;t) = |\Psi(x_1,x_2;t)|^2$ is analyzed. In a recent work \cite{2018-Dobrzyniecki-PRA}, we have conducted an analysis along these lines for a two-boson system with contact interactions. It was shown that the dynamical properties depend significantly on the strength $g$ of interparticle interactions. As $g$ is tuned from repulsive to strongly attractive values, the dynamics undergoes a transition between two regimes: the first one is dominated by sequential tunneling, so that both bosons leave the well one after the other, while the second one is almost completely dominated by pair tunneling. Here we analyze how these results apply to systems with non-zero-range interactions by studying the evolution of the two-particle density profile $\rho_2(x_1,x_2;t)$. To more easily tell apart the distinct tunneling processes in our analysis, we divide the configuration space into three regions $\mathbf{P}_i$: 
\begin{equation}
\label{eq:regions}
\begin{aligned}
\mathbf{P}_2 &= \{(x_1,x_2) : x_1 \le x_\mathrm{B} \land x_2 \le x_\mathrm{B}\},\\
\mathbf{P}_1 &= \{(x_1,x_2) : (x_1 > x_\mathrm{B} \land x_2 \le x_\mathrm{B}) \\
\phantom{\mathbf{P}_1 }&\phantom{= \{(x_1,x_2) } \lor (x_1 \le x_\mathrm{B} \land x_2 > x_\mathrm{B})\},\\
\mathbf{P}_0 &= \{(x_1,x_2) : x_1 > x_\mathrm{B} \land x_2 > x_\mathrm{B}\},
\end{aligned}
\end{equation}
where $x_\mathrm{B} \approx \sqrt{2\lambda}$ is the position of the well boundary. The regions $\mathbf{P}_2,\mathbf{P}_1,\mathbf{P}_0$ encompass configurations with exactly two, one, or zero particles inside the well, respectively.

\subsection{Two-boson dynamics}

\begin{figure}[t]
\centering
 \includegraphics[width=\linewidth]{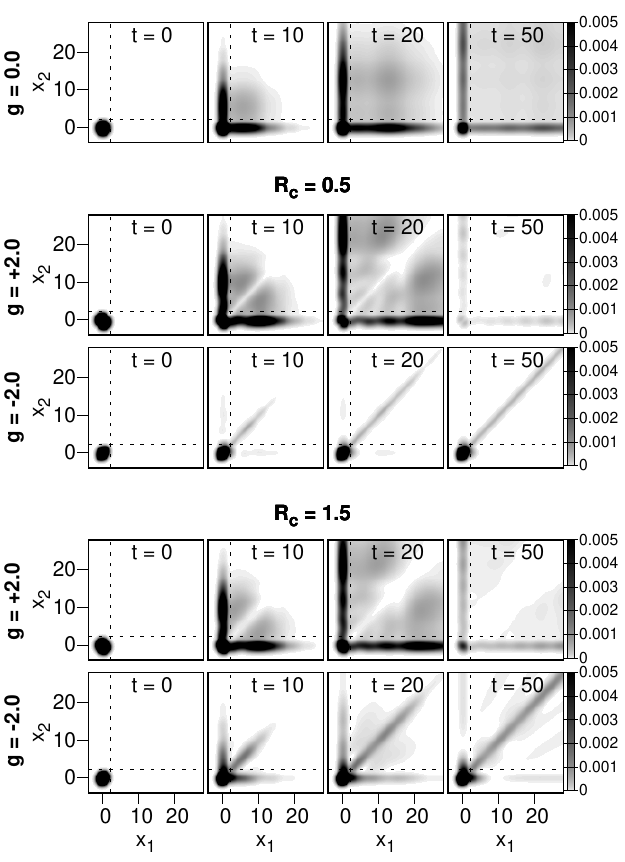}
 \caption{Time evolution of the density distribution $\rho_2(x_1,x_2,t)$ in an initially trapped two-boson system, for different interaction strengths $g$ and two different interaction ranges $R_\mathrm{c}$. The dashed lines demarcate the well boundary $x_\mathrm{B}\approx \sqrt{3}$. For the non-interacting and repulsive systems ($g = 0$, $g = 2$) essentially the entire decay process takes place via sequential tunneling of the two bosons. In the strongly attractive system ($g = -2$) the system decays mostly via pair tunneling, with the participation of sequential tunneling depending on interaction range $R_\mathrm{c}$. Positions and interaction range are in units of $\sqrt{\hbar/m\omega}$, interaction strength in units of $\sqrt{\hbar^3\omega/m}$, time in units of $1/\omega$.}
 \label{fig:BosonR-DensityEvolution}
\end{figure}

In Fig.~\ref{fig:BosonR-DensityEvolution} we show snapshots of the evolution of $\rho_2(x_1,x_2;t)$ at different times $t$ after opening the well, for two-boson systems with different interaction strengths $g$ and interaction ranges $R_\mathrm{c}$. For better visibility, the well boundary $x_\mathrm{B}\approx\sqrt{2\lambda}$ is indicated with dashed lines, dividing the configuration space into the different regions $\mathbf{P}_n$. At the beginning ($t = 0$), the entire two-body wave function is contained within the region $\mathbf{P}_2$. 

For the non-interacting system $(g = 0)$, both bosons tunnel entirely independently. After a short time $t = 10$ a large amount of density is present in the region $\mathbf{P}_1$, indicating a high probability of exactly one boson being outside the well. Additionally, a non-negligible amount of density is present in the region $\mathbf{P}_0$, corresponding to the event of two bosons having tunneled out of the well. Throughout the entire evolution, the two-body density is completely uncorrelated, \emph{i.e.}, the two-body wave function is simply the product of two identical one-body wave functions. The bosons are likely to leave the well one after the other, but a concidental simultaneous tunneling of two bosons is also possible. 

For the repulsive system $(g = +2)$, the sequential tunneling of bosons is enhanced. In this case, there is a visible anticorrelation in the boson positions, so that density close to the $x_1 = x_2$ diagonal vanishes. The tunneling here occurs solely via sequential tunneling, so that the probability flows from $\mathbf{P}_2$ into the $\mathbf{P}_1$ region, and subsequently from the areas of increased density in $\mathbf{P}_1$ into $\mathbf{P}_0$ (corresponding to the escape of the second boson out of the well). The tunneling of bound boson pairs is entirely absent. This is expected, since we have already noted in chapter \ref{sec:eigenstate-spectrum} that no bound pair states are available (in the outside-well region) for $g \ge 0$. Comparing the $R_\mathrm{c} = 0.5$ and $R_\mathrm{c} = 1.5$ cases, we see that the density dynamics remain qualitatively unchanged upon tuning of $R_\mathrm{c}$. 

The dynamics are significantly different for a strongly attractive system $(g = -2)$. Here, bound pair states are available for bosons in open space, and so pair tunneling is possible. For the $R_\mathrm{c} = 0.5$ case, we see that pair tunneling is essentially the only tunneling mechanism available. Therefore, the density flows directly from $\mathbf{P}_2$ into the $\mathbf{P}_0$ region and remains concentrated along the $x_1 = x_2$ diagonal, while it practically vanishes in the region $\mathbf{P}_1$. This demonstrates that the bosonic system with nonzero interaction range can undergo a transition into the pair tunneling regime, similarly to a $\delta$ interaction system. 

However, for the same $g = -2$ but a larger interaction range $R_\mathrm{c} = 1.5$, the density dynamics change. While the majority of the decay still takes place through pair tunneling, there is also non-negligible participation from sequential tunneling, as seen by the flow of density into $\mathbf{P}_1$. This can be explained by considering the system energy. The suppression of sequential tunneling occurs when the total system energy $E_\mathrm{INI}$ falls below the threshold of one-particle energy \cite{2018-Dobrzyniecki-PRA}. Since for larger $R_\mathrm{c}$ the energy of the attractive two-boson system becomes less sensitive to $g$ (as we have shown in Fig.~\ref{fig:InitialEnergy}a), the energy is farther away from crossing the threshold and the sequential tunneling is not as heavily suppressed. This also indicates that the interaction range parameter $R_\mathrm{c}$ can be treated as an additional knob to control the nature of tunneling, in addition to the interaction strength $g$. 

\subsection{Two-fermion dynamics}

\begin{figure}[t]
\centering
 \includegraphics[width=\linewidth]{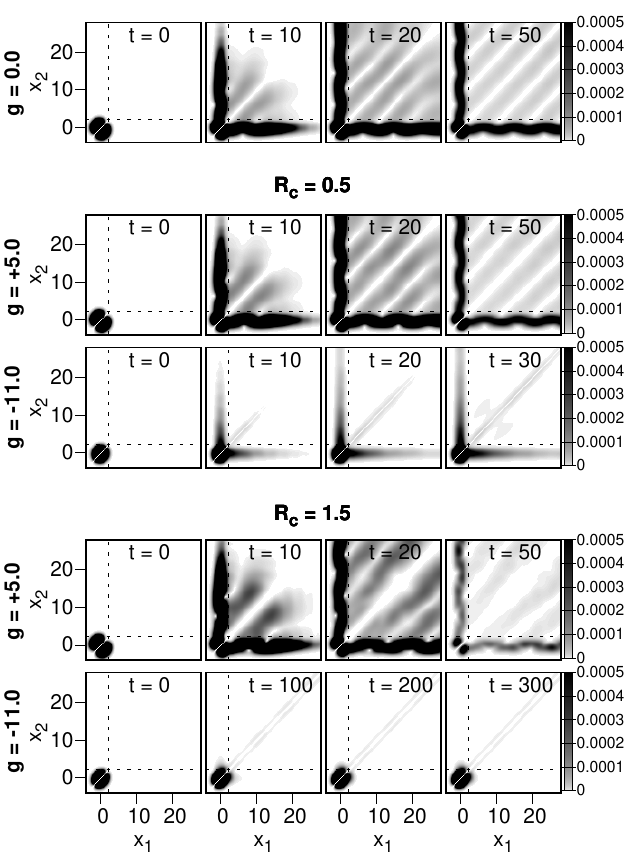}
 \caption{Time evolution of the density distribution $\rho_2(x_1,x_2,t)$ in an initially trapped two-fermion system, for different interaction strengths $g$ and two different interaction ranges $R_\mathrm{c}$. The dashed lines demarcate the well boundary $x_\mathrm{B}\approx \sqrt{5}$. For the non-interacting and repulsive systems ($g = 0$, $g = +5$) essentially the entire decay process takes place via sequential tunneling of the two fermions. In the strongly attractive system ($g = -11$) the system decays mostly via pair tunneling, with the participation of sequential tunneling depending on interaction range $R_\mathrm{c}$. Positions and interaction range are in units of $\sqrt{\hbar/m\omega}$, interaction strength in units of $\sqrt{\hbar^3\omega/m}$, time in units of $1/\omega$. }
 \label{fig:FermionR-DensityEvolution}
\end{figure}

We now proceed to analyze the density dynamics for a system of two fermions, and compare the result with the bosonic case. In Fig. \ref{fig:FermionR-DensityEvolution} we show the evolution of $\rho_2(x_1,x_2;t)$ for the two-fermion system at different interaction strengths $g$ and ranges $R_\mathrm{c}$. 

Already in the non-interacting case $(g = 0)$ the two-fermion dynamics differs significantly from the bosonic case. Now, the two-body wave function is no longer a product of two identical one-body wave functions. As a result, nonzero interparticle correlations are present in the system (although they are trivial, caused solely by the particle statistics). The density at the $x_1 = x_2$ diagonal remains zero for all times, and simultaneous tunneling of two fermions is suppressed. The only tunneling mechanism in this non-interacting case is the sequential tunneling, with density flowing from $\mathbf{P}_2$ to $\mathbf{P}_1$, and from there to $\mathbf{P}_0$. 

One characteristic feature is that, after a brief time, series of stripes of zero density appear in the $\mathbf{P}_0$ region, parallel to the $x_1 = x_2$ diagonal. Their presence can be simply explained as a result of interference between the wave functions of two approximately free particles with different momenta. In this approximation, the two-body density in the $\mathbf{P}_0$ region takes the form $\rho_2(x_1,x_2) \approx |e^{i k_1 x_1} e^{i k_2 x_2} - e^{i k_2 x_1} e^{i k_1 x_2}|^2 = 2[1-\cos[(k_2-k_1)(x_1-x_2)]]$, reproducing the interference pattern. If the momenta are chosen as $k_1 = 1,k_2 = \sqrt{3}$ (to match the initial fermion energies $E=1/2, E=3/2$),  this approximate form closely reproduces the observed spacing between the stripes. 

Now let us look at the fermion density dynamics in the case of repulsive interactions $(g = +5)$. For a relatively small interaction range $(R_\mathrm{c} = 0.5)$, since the density is already nearly zero close to the $x_1=x_2$ diagonal, the dynamics remain nearly unchanged from the non-interacting case. However, for a larger range $R_\mathrm{c} = 1.5$, the interactions are able to affect the dynamics significantly. In particular, there is a visible change in the shape of the interference minima within $\mathbf{P}_0$. 

We now turn to a case of the strongly attractive system $(g = -11)$. At this value of $g$ a fermionic pair mode is available, and the initial state can decay via pair tunneling. For the $R_\mathrm{c} = 0.5$ case, the pair tunneling is seen as an area of high density concentrated along the $x_1=x_2$ diagonal. However, sequential tunneling still plays a significant role, as indicated by the flow of density from $\mathbf{P}_2$ into $\mathbf{P}_1$. For $R_\mathrm{c} = 1.5$, however, sequential tunneling vanishes and fermions are only emitted as pairs. Thus, we see that for fermions, there exists a regime dominated by pair tunneling just like for bosons. Note also that the influence of $R_\mathrm{c}$ on the dynamics is quite opposite than in the bosonic case: increasing $R_\mathrm{c}$ causes a greater suppression of sequential tunneling. This effect is consistent with the total energy of the system. As we have seen in Fig.~\ref{fig:InitialEnergy}c, the energy $E_\mathrm{INI}$ becomes smaller upon increasing the interaction range to $R_\mathrm{c}=1.5$, thus it crosses the critical threshold of one-particle energy and one-body tunneling is suppressed more heavily. 

\section{Long-time dynamics and the decay rate}
\label{sec:long-time-dynamics}

The short-time dynamics, expressed through the evolution of $\rho_2$, allow us to distinguish between specifical tunneling mechanisms. However, a more in-depth understanding of the tunneling process can be gained by simulating the time evolution over longer timescales. In this chapter we will focus on long-time dynamics of the system, and in particular on the exponential nature of the decay which becomes evident at such timescales.

\begin{figure}[t]
\centering
 \includegraphics[width=\linewidth]{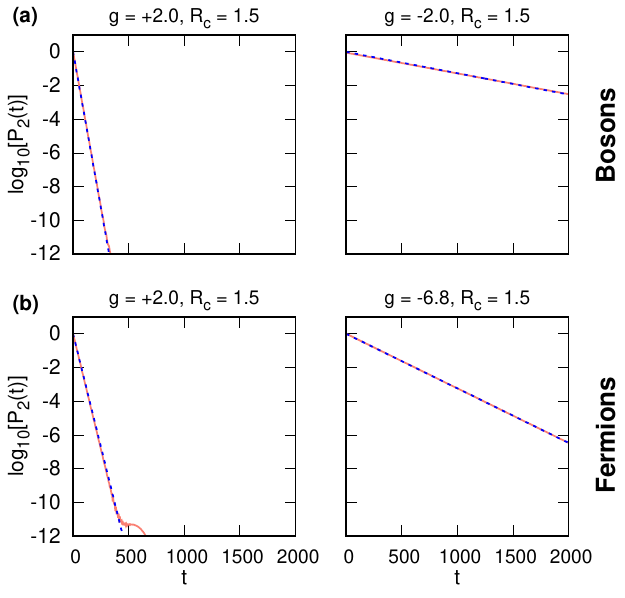}
 \caption{The time evolution of the probability ${\cal P}_2(t)$ over a long time scale (red, solid) for the two-particle system with various interaction strengths $g$, for the two-boson and two-fermion system with interaction range $R_\mathrm{c} = 1.5$. Blue dashed line shows an exponential fit to ${\cal P}_2(t)$. It can be seen that ${\cal P}_2(t)$ decays exponentially (apart from very long times). Time is given in units of $1/\omega$, interaction strength $g$ in units of $\sqrt{\hbar^3\omega/m}$, interaction range $R_\mathrm{c}$ in units of $\sqrt{\hbar/m\omega}$.}
 \label{fig:Rydberg-P2ExponentialDecay}
\end{figure}

It is known that decaying systems typically obey an exponential decay law \cite{1976-Davydov-Book}. That is, the survival probability, \emph{i.e.}, the probability that the system remains in the initial state, obeys an exponential decay law to a very good approximation (apart from very short and very long times \cite{1958-Khalfin-JETP,1997-Wilkinson-Nature,1972-Fonda-NuovoCim,2002-VanDijk-PRC,2006-Rothe-PRL,2006-Muga-PRA}). For the two-body trapped system, the survival probability is closely mimicked by the probability that both particles remain in the well region, given by $\mathcal{P}_2(t) = \int_{\mathbf{P}_2} |\Psi(x_1,x_2;t)|^2 \mathrm{d}x_1 \mathrm{d}x_2$. Therefore, its time evolution should be approximately given by 
\begin{equation}
\label{eq:exponential_decay_of_p2}
\mathcal{P}_2(t) \sim e^{-\gamma t},
\end{equation}
with the decay rate $\gamma$ constant in time. To confirm this assumption, in Fig.~\ref{fig:Rydberg-P2ExponentialDecay} we show the long-time evolution of $\mathcal{P}_2(t)$ for various interaction strengths, for bosons and fermions with interaction range $R_\mathrm{c}=1.5$. We compare the results to a fitted exponential function \eqref{eq:exponential_decay_of_p2}. The obtained decay rate $\gamma$ depends essentially on the interaction parameters. It is seen that $\mathcal{P}_2(t)$ indeed decays exponentially throughout nearly the entire evolution, regardless of $g$, both for bosons and fermions. Any deviations from exponential decay only occur at very short times, or at long times where the trapped system is practicaly completely depleted and $\mathcal{P}_2(t)$ is negligible. The decay rate $\gamma$ can be therefore determined by measuring the evolution of $\mathcal{P}_2(t)$ in time, and then fitting an exponential function to the results. In this way, the decay process for any value of $g$ and $R_\mathrm{c}$ can be characterized by a single value $\gamma$. At this point we wish to emphasize that the results presented for $\mathcal{P}_2$, in contrast to other probabilities, are almost insensitive to the details of the absorbing potential method used (for details, see appendix \ref{sec:cap-appendix}).

\subsection{Two-boson decay rate}

\begin{figure}[t]
\centering
 \includegraphics[width=1.0\linewidth]{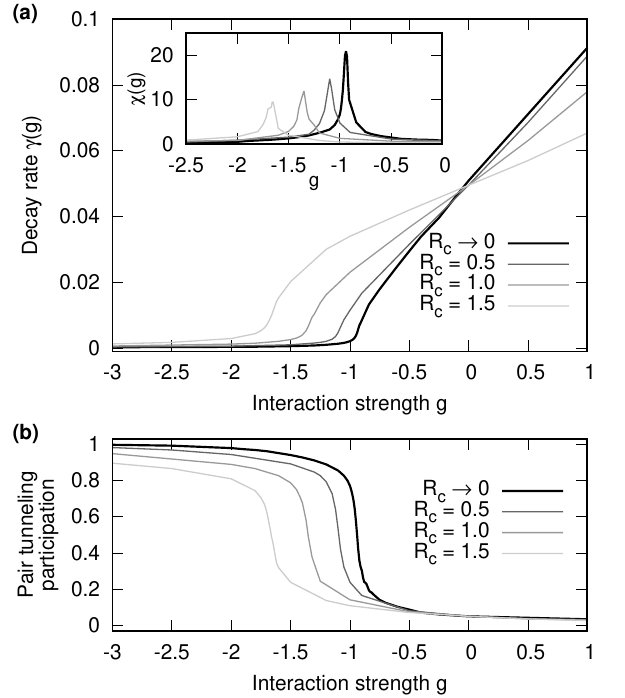}
 \caption{(a) The decay rate $\gamma(g)$ as a function of $g$ and $R_\mathrm{c}$, for the two-boson system. (Inset) The susceptibility $\chi(g) = \gamma^{-1} (\partial \gamma / \partial g)$. (b) The ratio $J_0/J$, expressing the relative participation of pair tunneling in the overall tunneling dynamics of the two-boson system. Interaction strength $g$ is expressed in units of $\sqrt{\hbar^3\omega/m}$, interaction range in units of $\sqrt{\hbar/m\omega}$, decay rate in units of $\omega$, susceptibility in units of $\sqrt{m/\hbar^3\omega}$.}
 \label{fig:BosonR-DecayRate}
\end{figure}

For a two-boson system, the obtained decay rate is shown in Fig.~\ref{fig:BosonR-DecayRate}a as a function of $g$, for different interaction ranges $R_\mathrm{c}$. We also include results in the contact interaction limit $R_\mathrm{c} \rightarrow 0$, \emph{i.e.}, for bosons interacting via the potential $g \delta(r)$. In the inset, we additionally show the susceptibility $\chi(g) = \gamma^{-1}(\partial \gamma/\partial g)$. Its peaks signal a large sensitivity of the decay rate to small changes of the interaction strength. 

It is seen that in the $R_\mathrm{c} \rightarrow 0$ limit, the decay rate $\gamma(g)$ displays a characteristic change in behavior approximately around the critical interaction strength $g_0 \approx -0.9$, so that the growth of $\gamma(g)$ is a lot faster above this point than below it, and a peak appears in $\chi(g)$. This change in behavior of $\gamma(g)$ is associated with the switch to the regime dominated by pair tunneling. Below the critical interaction strength $g_0$, sequential tunneling is suppressed, and the much slower pair tunneling is almost the only available decay mechanism \cite{2018-Dobrzyniecki-PRA}. 

As $R_\mathrm{c}$ increases from zero, the characteristic shape of $\gamma(g)$ and $\chi(g)$ is preserved (including the transition at some specific point $g_0$), but the sensitivity of the decay rate to the interactions is modified. Specifically, as $R_\mathrm{c}$ increases, the decay rate becomes less sensitive to a change in interaction strength. In the $R_\mathrm{c} \to \infty$ limit, the interaction is not felt at all and $\gamma(g)$ is interaction independent.

The critical value $g_0$ is dependent on $R_\mathrm{c}$, and it moves towards stronger attractive interactions as $R_\mathrm{c}$ increases. This effect can likewise be treated as a reflection of an analogous behavior of the total system energy. As explained previously, $g_0$ is approximately equal to the interaction strength for which $E_\mathrm{INI}(g_0)$ equals the energy of a single trapped particle, $E_\mathrm{INI}(g_0) = 0.5$. The energy $E_\mathrm{INI}(g,R_\mathrm{c})$ becomes less sensitive to $g$ as $R_\mathrm{c}$ increases, and so lowering $E_\mathrm{INI}$ below this energy threshold requires stronger attractive interactions. In the $R_\mathrm{c} \to \infty$ limit, the interactions are not felt at all and thus $g_0$ approaches minus infinity. 

To show that the $g_0$ indeed corresponds to a transition between two different dynamical regimes, we calculate the relative participation of the two decay mechanisms (pair and sequential tunneling) by theoretical calculation of different probability fluxes through the potential barrier (for details of this procedure, see \cite{2018-Dobrzyniecki-PRA}). This participation is expressed by the magnitude $J_0/J$, where $J$ is the total probability flux going out of the region $\mathbf{P}_2$, and $J_0$ is the total flux going directly from region $\mathbf{P}_2$ into $\mathbf{P}_0$. Therefore, the ratio $J_0/J$ expresses the relative probability that the initial state will decay by the emission of a bound pair, as opposed to a single boson. In Fig.~\ref{fig:BosonR-DecayRate}b we show the relative participation $J_0/J$ as a function of $g$, for different $R_\mathrm{c}$. It can be seen that an abrupt transition between two regimes indeed occurs at $g_0$. For $g > g_0$ the participation $J_0/J$ is near zero, indicating that nearly the entire tunneling takes place via sequential tunneling. For $g < g_0$, on the other hand, $J_0/J$ is close to one, indicating a near-total dominance of pair tunneling. 

It should be noted that the above analysis has significance for experimental practice, since it points to a method of achieving a more complete experimental control over the properties of the tunneling system. Specifically, by regulating the parameter $R_\mathrm{c}$ one can regulate the value of the critical interaction strength $g_0$ where the tunneling mechanism dominance is changed. Conversely, experimentally finding the value of $g_0$ can help in determining the effective interaction range $R_\mathrm{c}$. 

\subsection{Two-fermion decay rate}

\begin{figure}[t]
\centering
 \includegraphics[width=1.0\linewidth]{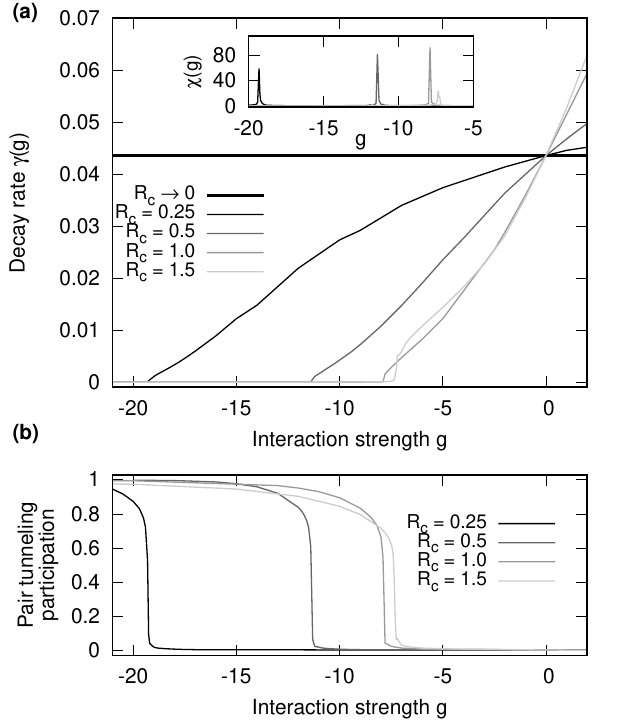}
 \caption{(a) The decay rate $\gamma(g)$ as a function of $g$ and $R_\mathrm{c}$, for the two-fermion system. (Inset) The susceptibility $\chi(g) = \gamma^{-1} (\partial \gamma / \partial g)$. (b) The ratio $J_0/J$, expressing the relative participation of pair tunneling in the overall tunneling dynamics of the two-fermion system. Interaction strength $g$ is expressed in units of $\sqrt{\hbar^3\omega/m}$, interaction range in units of $\sqrt{\hbar/m\omega}$, decay rate in units of $\omega$, susceptibility in units of $\sqrt{m/\hbar^3\omega}$.}
 \label{fig:FermionR-DecayRates}
\end{figure}

Now let us compare the above results with the case of the two-fermion system. In Fig.~\ref{fig:FermionR-DecayRates}a, we show the determined values of decay rate $\gamma(g)$ and its susceptibility $\chi(g)$ as a function of $g$, for different interaction ranges $R_\mathrm{c}$. It is seen that the decay rate behaves much the same as in the boson case, and in particular it is possible to identify an interaction strength $g_0$ at which the behavior of $\gamma(g)$ changes abruptly and a maximum appears in $\chi(g)$. The value of $g_0$, as in the bosonic case, can be approximately determined as the interaction strength for which the total energy of the system equals the one-particle energy, $E_\mathrm{INI}(g_0) = 0.5$. 

Analogously to the two-boson system, the dependence of $\gamma$ on $R_\mathrm{c}$ mimics the previously observed behavior of the initial energy $E_\mathrm{INI}(g)$ for two fermions. Thus, in the contact interaction limit ($R_\mathrm{c} \to 0$), the decay rate becomes independent of $g$ as the interactions vanish for fermionic atoms. For increasing $R_\mathrm{c}$, the sensitivity of $\gamma$ to a change of the interaction strength $g$ grows, quite opposite to the two-boson case. It should be noted, however, that this trend applies only to fairly small interaction ranges $R_\mathrm{c} \lesssim 1.0$. In the limit $R_\mathrm{c} \to \infty$ the decay rate approaches a constant, just as in the $R_\mathrm{c} \to 0$ case, for the same reason as in the bosonic case. Thus, for very high $R_\mathrm{c}$ (which are outside the scope of our work) the trend of increasing sensitivity is predicted to reverse. For example, it can be seen that at $R_\mathrm{c} = 1.5$ the slope of $\gamma(g)$ does not increase further, but is very close to that of $R_\mathrm{c} = 1.0$. 

In Fig.~\ref{fig:FermionR-DecayRates}b we show the pair tunneling participation $J_0/J$ as a function of $g$ for different interaction ranges $R_\mathrm{c}$. By comparing the figure with Fig.~\ref{fig:BosonR-DecayRate}b we see that, similarly to bosons, the interaction strength $g_0$ corresponds to a rapid transition between regimes dominated by pair and sequential tunneling. For small $R_\mathrm{c}$, increasing the interaction range causes the value of $g_0$ to move towards weaker attractions, which is quite opposite to the behavior of bosonic systems. However, also this trend is expected to reverse for larger interaction ranges. In the limit $R_\mathrm{c} \rightarrow 0$ fermion pairing and pair tunneling vanishes completely, and so in this limit the value of $g_0$ approaches minus infinity, just like in the bosonic case. 

It is also worth noting that the magnitude of $g_0$ for the two-fermion system is significantly larger than in the bosonic case. To illustrate this, one may consider that, while for the bosonic case changing the interaction range from $R_\mathrm{c}=0.5$ to $R_\mathrm{c} = 1.5$ causes a relatively small shift of $g_0$, for the fermionic case the analogous shift of $g_0$ is an order of magnitude greater. This can be explained by the fact that fermions in general feel the interactions less strongly than bosons, and the fact that, for fermions, pair tunneling doesn't appear at all until the interaction strength $g$ is below a value $g_\mathrm{pair} \sim -R_\mathrm{c}^{-1}$. 

\begin{figure}[t]
\centering
  \includegraphics[width=1.0\linewidth]{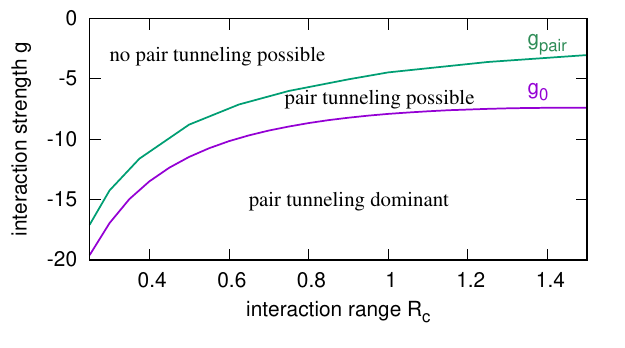}
 \caption{The value of two critical interaction strengths of the two-fermion system, as a function of $R_\mathrm{c}$: the value of $g_\mathrm{pair}$, below which fermions can form bound pairs, and the value $g_0$, below which the system transitions to a regime dominated by pair tunneling. Interaction strength is expressed in units of $\sqrt{\hbar^3\omega/m}$, interaction range in units of $\sqrt{\hbar/m\omega}$.}
 \label{fig:FermionR-CriticalInteractionStrength}
\end{figure}

Similarly as in the bosonic case, by tuning $R_\mathrm{c}$ one can manipulate the value of $g_0$. Furthermore, as the value $g_\mathrm{pair}$ is also dependent on $R_\mathrm{c}$, it can be treated as a second tunable parameter. In Fig.~\ref{fig:FermionR-CriticalInteractionStrength} we show $g_\mathrm{pair}$ and $g_0$ as a function of $R_\mathrm{c}$. It is seen that they can be regulated fairly extensively (though not independently) by tuning the interaction range $R_\mathrm{c}$, opening new ways to experimental control over the system properties. Conversely, one can attempt to find $g_\mathrm{pair}$ and/or $g_0$ to determine the value of $R_\mathrm{c}$. It is also worth noting that the value of the transition interaction strength $g_0$ can be to some degree manipulated by changing the shape of the potential outside of the well \cite{2018-Dobrzyniecki-PRA}, which suggests an additional way to change $g_0$ and $g_\mathrm{pair}$ independently from each other. 

\section{Conclusion}
\label{sec:conclusion}

We have examined the dynamical properties of a system of two Rydberg-dressed bosons or fermions with finite-range interactions, tunneling from a leaky potential well into the open space. The nature of the tunneling dynamics is found to depend significantly on the interaction strength. For the system with repulsive interactions, only sequential tunneling of two particles is available, independently of quantum statistics. For attractive interactions, the tunneling significantly depends on statistics. In the case of bosons, pair tunneling can be observed at any strength of the attractive interactions. In the case of fermions, pair tunneling can occur only for sufficiently strong attractive interactions $g < g_\mathrm{pair}$, with $g_\mathrm{pair}$ dependent on interaction range.

The proportional participation of pair tunneling in the overall tunneling process depends on the strength of the attractive interaction. We find that the dominant decay mechanism changes abruptly as the interaction strength crosses a critical value $g_0$. For weaker attractions $(g > g_0)$, the decay process occurs mainly by the sequential emission of two particles from the well. For stronger attractions $(g < g_0)$, sequential tunneling is suppressed, and the particles tunnel mainly as bound pairs. This transition occurs in a similar way both for bosonic and fermionic systems. However, the evolution of two-particle density correlations shows visible differences between the two cases. 

The interaction strengths required to reach the regime of dominant pair tunneling are found to be significantly different for bosons and fermions. For fermions, a much greater strength $|g|$ of attractive interactions is needed. This is both due to the vanishing of the fermion wave function at $x_1=x_2$, which weakens the influence of attractive interactions, and the fact that antisymmetric bound pair states become available only for $g < g_\mathrm{pair} < 0$, with $|g_\mathrm{pair}|$ having particularly large values at short interaction ranges.

Changing the interaction range $R_\mathrm{c}$ affects the decay rate of the system and the participation of the different decay mechanisms, in a quite opposite way for bosons and fermions. For bosons, increasing $R_\mathrm{c}$ (with interaction strength fixed) diminishes the effect of the interactions, so that the decay rate approaches the value of the non-interacting system. Also, the pair tunneling induced by attractive interactions becomes less dominant. For fermions, the situation is different because in the limit of zero-range interactions ($R_\mathrm{c} \to 0$) the interaction vanishes completely. As a result, in this case increasing $R_\mathrm{c}$ instead enhances the effect of the interactions. However, in the limit of very large interaction ranges, the decay rate approaches the non-interacting value both for bosons and fermions.

Compared to results for a system of contact-interacting bosons \cite{2018-Dobrzyniecki-PRA}, the tunneling of atoms with long-range interactions remains qualitatively similar. However, the use of longer-range interactions allows to extend the investigation to systems of identical fermions as well. The longer-range interactions also give an additional dimension of control over the system properties. Specifically, the critical values of the interaction strength, $g_0$ (below which pair tunneling becomes dominant) and $g_\mathrm{pair}$ (below which bound fermion pairs can appear), are dependent on the interaction range $R_\mathrm{c}$. This indicates that the interaction range $R_\mathrm{c}$ can be treated as an additional tunable parameter to exercise more complete control over the system properties. In light of the recent experiments with few-body tunneling systems \cite{2012-Zurn-PRL,2013-Zurn-PRL} and Rydberg-dressed atoms \cite{2016-Jau-NatPhys,2016-Zeiher-NatPhys,2017-Zeiher-PRX,2019-Arias-PRL,2020-Borish-PRL}, the results presented in this paper have potential significance for future research in this direction. 

Finally, it should be noted that in this work we have deliberately limited ourselves to interaction ranges $R_\mathrm{c} \le 1.5$, on the scale of the extension of the initial wave function. For higher ranges, the existing model may break down and the dynamical properties might become more complicated. In particular, the decay of the system may become significantly non-exponential in such cases, necessitating a new approach to analysis. 

\section{Acknowledgments}

This work was supported by the (Polish) National Science Center Grant No. 2016/22/E/ST2/00555. 

\appendix

\section{Application of the complex absorbing potential}
\label{sec:cap-appendix}

\begin{figure}[t]
\centering
  \includegraphics[width=\linewidth]{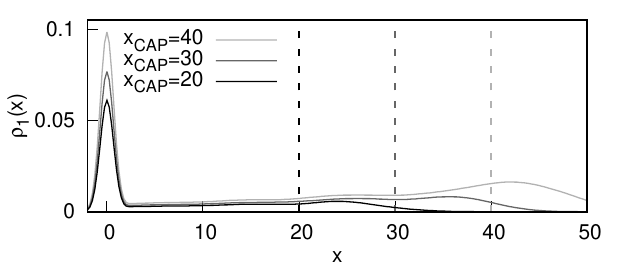}
 \caption{The one-body density distribution $\rho_1(x)$ of a tunneling two-boson system with $g = -0.5, R_c = 1.5$ at a specific time $t = 70$, for different values of the CAP position $x_\mathrm{CAP}$. Dashed lines indicate the positions $x_\mathrm{CAP}$ at which the absorbing potential begins. It can be seen that the wave function for $x > x_\mathrm{CAP}$ is gradually absorbed, so that the particles escaping from the well do not reach the simulation region boundary at $x=60$. However, the one-body density in the well region around $x = 0$ is also visibly affected by changing the CAP parameters, which indicates a limitation of the approach. Length is expressed in units of $\sqrt{\hbar/m \omega}$.}
 \label{fig:CAP-SingleParticleDensity}
\end{figure}

In order to simulate the infinite extent of space within the finite simulated domain $x \in [-4,60]$, we use the complex absorbing potential (CAP) technique. Namely, we add an imaginary potential term $-i \Gamma(x)$ to the single-particle Hamiltonian. Here $\Gamma(x)$ is chosen as a function which is zero in the region $x < x_\mathrm{CAP}$, and has a smoothly rising form $\alpha (x - x_\mathrm{CAP}) ^ \beta$ for $x \ge x_\mathrm{CAP}$. Throughout the work we use the parameter values $\alpha = 0.001, \beta = 2, x_\mathrm{CAP} = 30$. To ensure numerical accuracy, it is necessary to verify that the simulation results remain insensitive to changes to these parameters. 

\begin{figure}[t]
\centering
  \includegraphics[width=\linewidth]{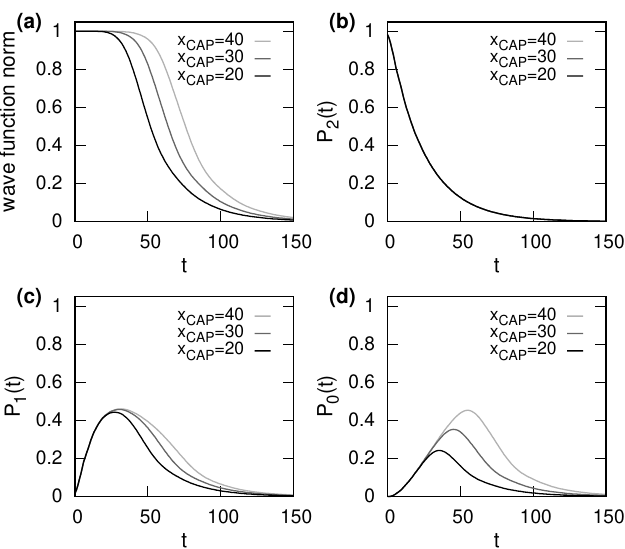}
 \caption{(a) Time evolution of the norm of the wave function $\mathcal{P}(t) = \int \rho_2(x_1,x_2;t) \mathrm{d}x_1 \mathrm{d}x_2$ for a two-boson system with $g = -0.5, R_c = 1.5$. Results are shown for different values of the CAP position $x_\mathrm{CAP}$. The norm diminishes over time as parts of the wave function are absorbed. (b,c,d) Time evolution of the partial probabilities $\mathcal{P}_n(t)$ ($n = 2,1,0$), i.e. the total probabilities in the regions $\mathbf{P}_n$, for different $x_\mathrm{CAP}$. For the probability $\mathcal{P}_2(t)$ the evolution remains insensitive to the CAP parameters (since it is only dependent on the state of the particles far from the CAP region). The probabilities $\mathcal{P}_1(t)$ and $\mathcal{P}_0(t)$, on the other hand, are significantly affected. Time is expressed in units of $1/\omega$, length in units of $\sqrt{\hbar/m \omega}$.}
 \label{fig:CAP-Norm}
\end{figure}

First let us demonstrate the effects of the absorbing potential. In Fig.~\ref{fig:CAP-SingleParticleDensity} we show a snapshot at time $t=70$ of the one-body density $\rho_1(x;t)$ for an example two-boson system, obtained for three different values of the CAP position $x_\mathrm{CAP}$. 
It can be seen that the parts of the wave function which overlap with the CAP region (delimited by the dashed lines) are gradually absorbed during the time evolution, which prevents unwanted reflections off the domain wall at $x = 60$. 
As a result, the evolution is non-unitary and the norm of the wave function $\mathcal{P}(t) = \int |\Psi(x_1,x_2;t)|^2 \mathrm{d}x_1 \mathrm{d}x_2$ is no longer conserved in time. This is shown in Fig.~\ref{fig:CAP-Norm}a, where we show the time evolution of $\mathcal{P}(t)$ for the example two-boson system. For short times the norm of the wave function is equal to unity, but it begins decreasing as soon as the tunneling particles start entering the CAP region, which occurs at an earlier time if $x_\mathrm{CAP}$ is placed closer to the well. 

The CAP technique allows to properly simulate unbounded systems. However, the use of complex absorbing potentials for few-body systems requires caution, since quantities that depend on the microscopic state of the system in the absorption region can be strongly affected by this artificial mechanism. For example, in Fig.~\ref{fig:CAP-SingleParticleDensity} one can see that the one-body density in the well region becomes different when the simulation is performed with different CAP parameters. However, quantities that depend \emph{only} on the two-body wave function $\Psi(x_1,x_2)$ calculated far away from the CAP ($x_1, x_2 \ll x_\mathrm{CAP}$) are captured properly. To demonstrate this, we divide the total probability $\mathcal{P}(t)$ into partial probabilities $\mathcal{P}_n(t) = \int_{\mathbf{P}_n} \rho_2(x_1,x_2;t) \mathrm{d}x_1 \mathrm{d}x_2$ ($n = 2,1,0$), where the regions $\mathbf{P}_n$ are defined as in \eqref{eq:regions}. Note that $\mathcal{P}_0(t)$ and $\mathcal{P}_1(t)$ are dependent on the state of the particles far from the well, while $\mathcal{P}_2(t)$ depends only on the state of particles close to the well. 
In Fig.~\ref{fig:CAP-Norm} we show the evolution of these probabilities for the example system of two bosons ($g = -0.5, R_c = 1.5$), with different values chosen for $x_\mathrm{CAP}$. 
It is clear that the obtained values of $\mathcal{P}_1(t)$ and $\mathcal{P}_0(t)$ (Fig.~\ref{fig:CAP-Norm}c,d) are highly sensitive to CAP parameters, indicating that they cannot be accurately predicted with this approach. On the other hand, the evolution of the probability $\mathcal{P}_2(t)$ (Fig.~\ref{fig:CAP-Norm}b) remains essentially unaffected by the absorbing potential. A similar analysis performed for fermionic systems leads to the same conclusions. Note that for larger particle numbers ($N > 2$), only the total $N$-particle density $\rho_N(x_1,\ldots,x_N;t) = |\Psi(x_1,\ldots,x_N;t)|^2$ is insensitive to the CAP approach (in the region where $x_1,\ldots,x_N \ll x_\mathrm{CAP}$).

\bibliography{_Biblio}  

\end{document}